\documentclass[12pt]{article}
\usepackage{graphicx}
\usepackage{color}
\usepackage{comment}
\usepackage{amsmath}
\usepackage{hyperref}

\def\Title#1{\begin{center} {\Large #1 } \end{center}}
\def\Author#1{\begin{center}{ \sc #1} \end{center}}
\def\Address#1{\begin{center}{ \it #1} \end{center}}

\newcommand\pubblock{\rightline{\begin{tabular}{l}\\
           \end{tabular}}}
\newenvironment{Abstract}{\begin{quotation}  }{\end{quotation}}
\newenvironment{Presented}{\begin{quotation} \begin{center} 
             PRESENTED AT\end{center}\bigskip 
      \begin{center}\begin{large}}{\end{large}\end{center} \end{quotation}}
\def\Acknowledgements{\bigskip  \bigskip \begin{center} \begin{large}
             \bf ACKNOWLEDGEMENTS \end{large}\end{center}}

\def\beq{\begin{equation}}
\def\eeq#1{\label{#1}\end{equation}}
\def\eeqn{\end{equation}}

\def\beqa{\begin{eqnarray}}
\def\eeqa#1{\label{#1}\end{eqnarray}}
\def\eeqan{\end{eqnarray}}

\let\bar=\overbar

\def\Dslash{\not{\hbox{\kern-4pt $D$}}}
\def\dslash{\not{\hbox{\kern-2pt $\del$}}}

\def\msb{{\bar{\ssstyle M \kern -1pt S}}}

\begin{document}
\begin{titlepage}
\pubblock

\vfill
\Title{Theia: A multi-purpose water-based liquid scintillator detector}
\vfill
\Author{Vincent Fischer on behalf of the Theia collaboration}
\Address{University of California at Davis, Department of Physics, Davis, CA 95616, U.S.A.}
\vfill
\begin{Abstract}
Recent developments in the field of liquid scintillator chemistry and fast-timing photosensors paved the way for a new generation of large-scale detectors capable of tackling a broad range of physics issues. 
Water-based Liquid Scintillator is a novel detection medium that combines the advantages of pure water, including low attenuation, accurate direction reconstruction, and low cost, and those of liquid scintillator, including high light yield and low-threshold detection.
When coupled with high efficiency, fast-timing photosensors, such as Large Area Picosecond PhotoDetectors, Water-based Liquid Scintillator exhibits an immense potential for neutrino physics and BSM searches. 
Theia is a 50-kiloton multi-purpose neutrino detector that aims to jointly deploy these two technologies in order to fulfill its physics program objectives, including the determination of the neutrino mass hierarchy and the CP violation phase in the leptonic sector, the detection of solar, reactor, and supernova neutrinos, and the search for neutrinoless double beta decay and proton decay.
\end{Abstract}
\vfill
\begin{Presented}
Thirteenth Conference on the Intersections of Particle And Nuclear Physics (CIPANP2018)\\
Palm Springs, U.S.A., May 29 -- June 3, 2018
\end{Presented}
\vfill
\end{titlepage}
\def\thefootnote{\fnsymbol{footnote}}
\setcounter{footnote}{0}
%

\section{Introduction}

Water and liquid scintillator have been widely utilized in neutrino physics for more than 50~years. 
Both detection media display different characteristics and, in most cases, the choice of using one or the other depends on the physics goals being pursued. 
While liquid scintillator detectors are more fit to detect low-energy events due to the high light yield of the scintillation process, water detectors are more fit to detect and reconstruct higher energy events thanks to the directional nature of Cherenkov light emission.
Recent R\&D in the field of liquid scintillator chemistry led to the development of a novel kind of detection medium: Water-based Liquid Scintillator (WbLS), a mixture of water and liquid scintillator~\cite{Yeh:2011zz}(see Section~\ref{sec:wbls}).

Large Area Picosecond PhotoDetectors (LAPPDs)\cite{Adams:2016tfm} are fast-timing photosensors based on the principle of MicroChannel plates (MCP) (see Section~\ref{sec:lappd}). 
Their excellent time and position resolution allows precise charged particle track reconstruction.

Theia is a future project that aims at using WbLS along with LAPPDs in a large scale multi-purpose detector~\cite{Alonso:2014fwf}.
The design of the Theia detector is discussed in Section~\ref{sec:detector} and its multiple physics goals are described in Section~\ref{sec:physics}.

\section{Water-based Liquid Scintillator}
\label{sec:wbls}

Developed and manufactured at Brookhaven National Laboratory, Water-based Liquid Scintillator is a mixture of pure water and oil-based liquid scintillator.
Since these liquids are not miscible, an additional compound, called a surfactant, is needed for the mixture to be homogeneous and stable over time.
Surfactants, such as PRS (Linear Alkyl Sulfonate), are substances that lower the surface tension between two liquids hence allowing them to mix.
When in contact with water and scintillator, the surfactant molecules, having an hydrophilic head and an hydrophobic tail, create structures called "micelles" in which scintillator molecules are held. 
The micelles being miscible with the aqueous solution, this allows WbLS to be an homogeneous mixture.

The nature of WbLS allows it to be tunable in terms of liquid scintillator content. When trying to study low energy physics, such as solar neutrinos or double-beta decay, the advantages of pure liquid scintillator - high light yield and low energy threshold - can be exploited by using WbLS with a high scintillator content (e.g. $>$20\%). 
On the other hand, higher energy physics, such as neutrino oscillations searches, can profit from the advantages brought by pure water - Cherenkov directionality, low light attenuation and low cost - in a WbLS mixture with a low scintillator content (e.g. $<$5\%).\\

Since isotope loading (gadolinium, lithium, tellurium, etc..) have been performed with both water and liquids scintillator for different purposes (neutron detection, charged current events detection, double-beta decay searches, etc...), WbLS could easily be loaded with a soluble metallic compound in order to increase its performances for a broader range of physics searches.\\

Studies are currently being performed to understand and assess the capabilities that fast-timing photosensors could bring in reconstructing charged particles tracks in WbLS through the separation of Cherenkov light (fast, directional emission) and scintillation light (slower, isotropic emission) using time and charge information~\cite{Caravaca:2016fjg}.\\

\section{Large Area Picosecond PhotoDetectors}
\label{sec:lappd}

\begin{figure}[htb]
\centering
\includegraphics[width=0.49\textwidth]{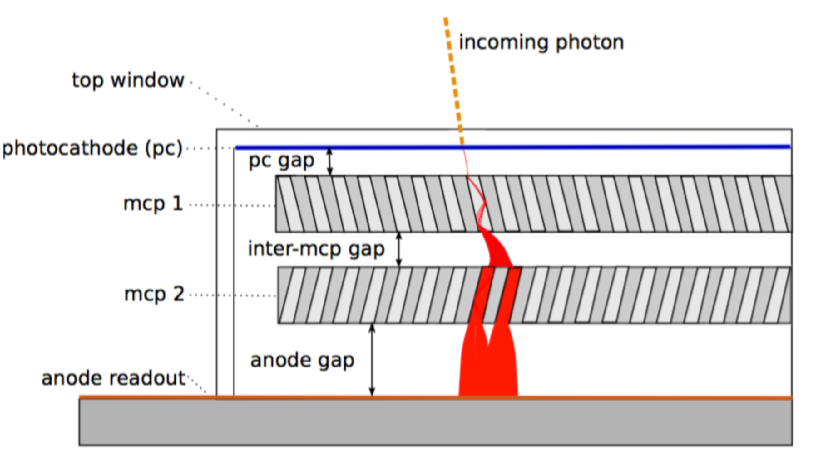}
\includegraphics[width=0.49\textwidth]{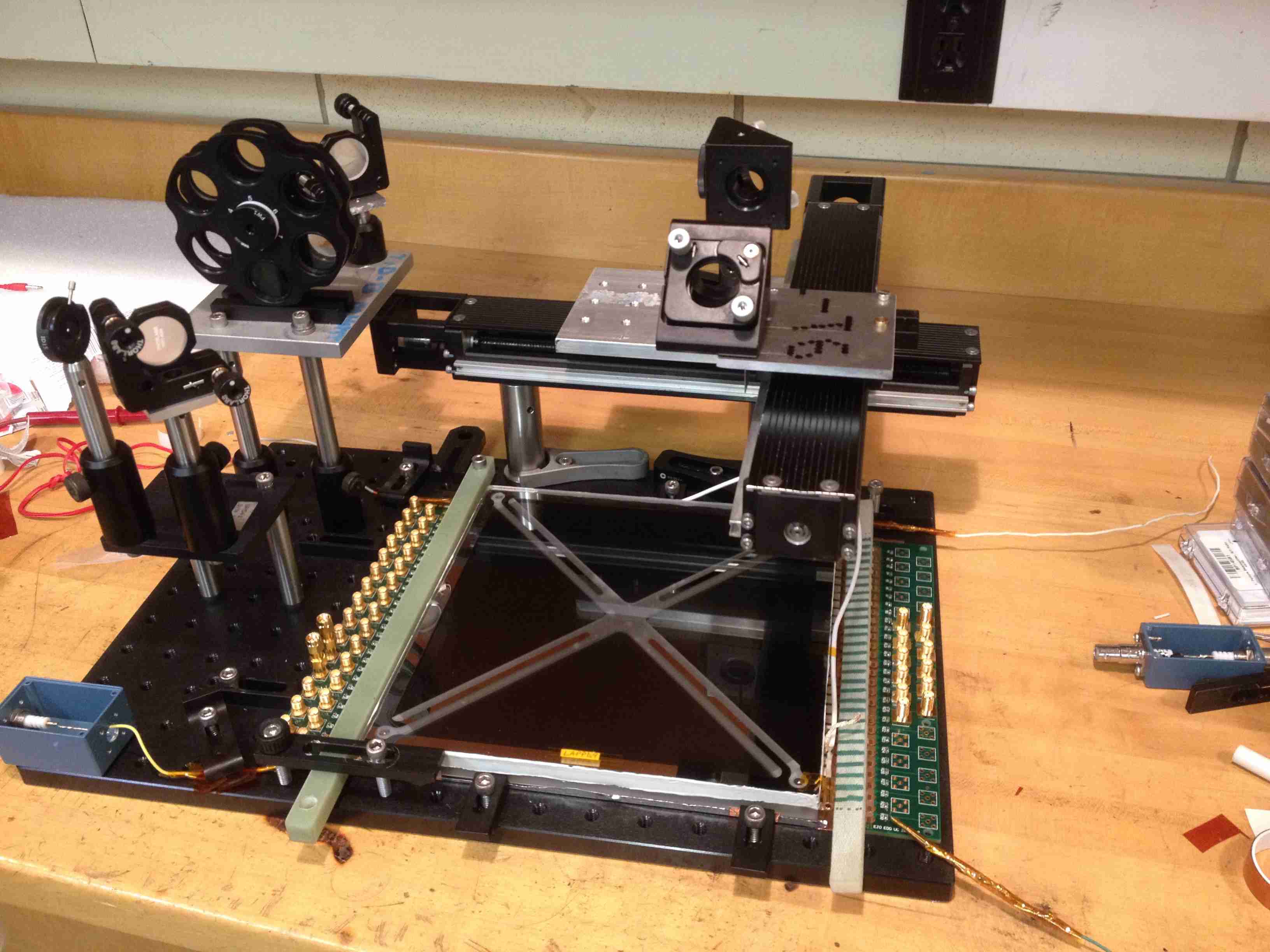}
\caption{Left: Detection principle of an MCP-based LAPPD. The incoming photon creates a photoelectron at the photocathode, which gets multiplied before being detected on the anode. Right: Picture of an LAPPD on a test-stand at Iowa State University (Courtesy of Matthew Wetstein). }
\label{fig:lappds_pic}
\end{figure}

LAPPDs are 20x20~cm MCP-based tiles (see Figure~\ref{fig:lappds_pic} (left)) with a single photoelectron time resolution of $\sim$50~ns, and a spatial resolution less than a centimeter. 

Such design allows LAPPDs to offer a significant advantage over conventional photomultiplier tubes (PMTs).
While PMTs are single-pixel detectors, LAPPDs can be considered imaging detectors able to reconstruct light patterns, such as Cherenkov rings, in both time and position.
Their spatial resolution allows precise neutrino interaction vertex reconstruction as well as separation of multiples rings originating from the same event, such as pion-induced tracks.\\

LAPPDs are currently developed and manufactured by the Incom company.
While some tiles have been tested individually on test-stands using LEDs or lasers, they haven't been deployed in an actual neutrino experiment.
The ANNIE experiment~\cite{Back:2017kfo} plans on operating at least 5~LAPPDs in their detector during the Phase~II of the experiment in order to improve the neutrino interaction vertex position reconstruction.
Several of the tiles expected to be installed in ANNIE are currently being thoroughly tested at Iowa State University (see Figure~\ref{fig:lappds_pic} (right)) in order to asses their capabilities in terms of gain, time resolution and noise levels.

\section{The Theia detector}
\label{sec:detector}

Theia being a future detector still being investigated, the following description is an initial design subject to changes.

The design of Theia is based on existing detectors such as Super Kamiokande~\cite{Fukuda:2002uc} and SNO+~\cite{Andringa:2015tza}. 
In order to limit cosmogenic backgrounds for rare events searches, Theia must be operated underground with an overburden larger than 4,000~meters water equivalent.
An outer shield, instrumented with PMTs, and filled with water will act as a barrier for external radioactivity originating from the surrounding rock.
The base design for the Theia target is a 50~kilotonnes volume instrumented with more than 100,000~photosensors (conventional PMTs and LAPPDs) in order to reach an effective photocoverage of more than 90\%.

The nominal detection medium for Theia is water-based liquid scintillator.
The concentration of scintillator is yet unknown and will depend on the desired sensitivity to some of the physics goals presented in Section~\ref{sec:physics}.
A possible loading of the liquid with a metallic compound is also being investigated as well as the possibility of deploying an inner bag filled with an isotope suited for double-beta searches - a concept similar to KamLAND-Zen~\cite{KamLAND-Zen:2016pfg}.


\section{The physics of Theia}
\label{sec:physics}

In the following section, some of the main physics goals of Theia will be presented: long baseline neutrino oscillations and double-beta decay searches.
However, Theia will be able to cover a much broader range of physics searches, including but not limited to: solar, reactor, supernova and geological neutrinos, nucleon decay searches, sterile neutrino oscillation searches, etc...

\subsection{Long baseline neutrino oscillations}
\label{subsec:osc}

\begin{figure}[htb]
\centering
\includegraphics[width=0.45\textwidth]{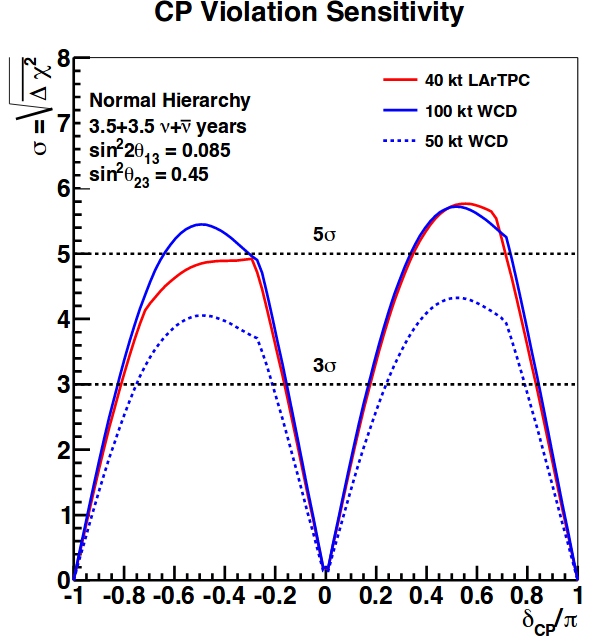}
\includegraphics[width=0.45\textwidth]{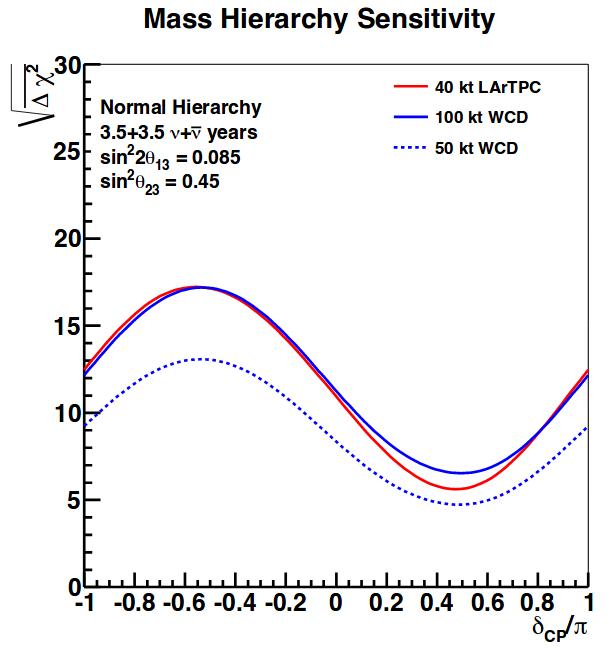}
\caption{Expected sensitivity to CP violation (left) and mass hierarchy (right) for a 40~kt liquid argon detector (DUNE) and a water Cherenkov detector (50 and 100~kt), both built at the Sanford Underground Research Facility and exposed to the future LBNF neutrino beam from Fermilab. Reaching those levels of sensitivity can be achieved if the neutral current background is reduced to 25\% of its value in SuperK-I.}
\label{fig:theia_osc}
\end{figure}

The possibility of using a water Cherenkov detector to perform long baseline neutrino oscillations using a beam from the Fermi National Accelerator Laboratory has been discussed in the past~\cite{Barger:2007yw}.
The current design of the experiment, now named DUNE (Deep Underground Neutrino Experiment)~\cite{Adams:2013qkq}, focuses on the use of liquid argon as a detection medium in its four 10-kt detectors.
Studies show that with advances to water Cherenkov reconstruction techniques, a 100-kt water Cherenkov detector could have comparable sensitivity to the observation of CP violation in the neutrino sector and the determination of the neutrino mass hierarchy, as shown in Figure~\ref{fig:theia_osc}.

The effects of having WbLS as a detection medium instead of water have not been taken into account yet in the previous studies.
However, the increase of light yield brought by scintillation is expected to increase the energy resolution of the detector while keeping the directional information extracted from the Cherenkov light at levels similar to those of pure water.

Besides, the low energy threshold will allow a WbLS-based detector such as Theia to detect neutrons created upon a neutrino interaction through radiative capture on hydrogen (or on loaded metals such as gadolinium).
Detecting those neutrons would help reduce systematic energy shifts occurring when mis-reconstructing those events as perfect charged-current quasi-elastic (CCQE) interactions~\cite{Martini:2012uc}.

Furthermore, the use of fast-timing photosensors such as LAPPDs would allow better track reconstruction capabilities than what is already achievable with conventional PMTs.\\

DUNE being already underway, Theia must not be seen as a competitor but as a complementary experiment. 
Installing Theia at SURF, next to DUNE, would provide the overburden needed for the experiment and would allow the two experiments, once combined, to reach sensitivities of more than 5$\sigma$ over half of the $\delta_{\text{CP}}$ range for CP violation discovery and over the entire $\delta_{\text{CP}}$ range for mass hierarchy determination.

\subsection{Neutrinoless double-beta decay}
\label{subsec:dbd}

Searching for neutrinoless double-beta decay requires loading an additional isotope in WbLS.
At the moment, two isotopes are under investigation: $^{136}$Xe, used in KamLAND-Zen, and $^{130}$Te, expected to be used in SNO+.

The detector configuration for double-beta decay searches being investigated consists of deploying a smaller inner balloon in the target volume and fill it with pure liquid scintillator (Linear Alkyl Benzene, called LAB) loaded with a higher concentration of double-beta isotope - an approach similar to KamLAND-Zen.

Backgrounds can be greatly reduced by applying fiducialization and tagging techniques, such as triple coincidence detection and event topology reconstruction. 
The latter is of a particular interest in the case of WbLS given the possibility to tune the scintillation content of the medium and hence optimize the signal to background ratio of Cherenkov light against scintillation light while keeping relatively high levels of light yield.

\begin{figure}[htb]
\centering
\includegraphics[width=0.8\textwidth]{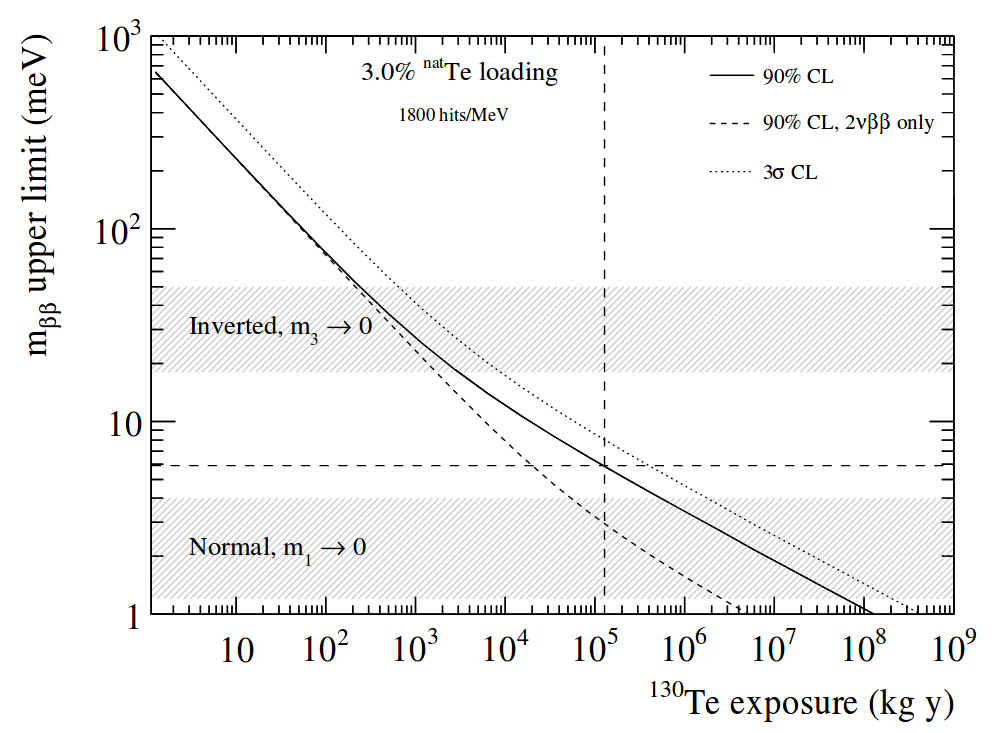}
\caption{Upper limit (90\%~CL) on the effective Majorana mass (m$_{\beta\beta}$, in meV) as a function of the exposure for a detector like Theia loaded with 3\% of natural tellurium by weight in a 8-meter balloon ($\sim$19 tonnes of $^{130}$Te). With an exposure of more than 10$^{5}$~kg.y, Theia is expecting to reach an upper limit on m$_{\beta\beta}$ of 6.3~meV and a limit on the neutrinoless double-beta decay half-life of $^{130}$Te of 1.1 $\times$ 10$^{28}$~years.}
\label{fig:theia_dbd}
\end{figure}

As shown in Figure~\ref{fig:theia_dbd}, given its large size, Theia will be able to completely probe the inverted hierarchy part of the mass phase space and reach limits on the effective Majorana mass m$_{\beta\beta}$ of less than 10~meV, corresponding to limits on the neutrinoless double-beta half-life of $^{130}$Te of more than 10$^{28}$~years.

The versatility of the loading technique paves the way for higher concentration of double-beta isotope in a possible second phase of the experiment, thus allowing Theia to start probing the normal hierarchy part of the Majorana mass phase space.


\section{Conclusions}

While Theia is still being considered a project for the time being, it draws a lot of interest from several institutions, including Fermilab.
A proto-collaboration has already been formed and includes more than 60 collaborators from 6 different countries.
The R\&D behind the design of Theia (novel kind of detection media and fast-timing photosensors) attracts a lot of groups already involved in experiments such as ANNIE, SNO+ and WATCHMAN~\cite{Askins:2015bmb}.

By combining the capabilities of WbLS and LAPPDs with the important size of Theia and the knowledge and expertise obtained from operating detectors such as ANNIE, SNO+ and WATCHMAN, one can easily expect Theia to be able to provide world-leading measurements for a broad range of physics measurements.


\Acknowledgements
The author is funded by the U.S. Department of Energy and the Nuclear Science and  Security  Consortium.

This  material  is  based  upon  work  supported  by  the  Department  of  Energy  National  Nuclear Security  Administration  through  the  Nuclear  Science and  Security  Consortium  under  Award Number DE-NA0003180.
This report was prepared as an account of work sponsored by an agency of the United States Government. Neither the United States Government nor any agency thereof, nor any of their employees,  makes any  warranty,  express  or  limited,  or  assumes  any  legal  liability  or responsibility  for  the  accuracy,  completeness,  or  usefulness  of  any  information,  apparatus, product,  or  process  disclosed,  or  represents  that  its  use  would  not  infringe  privately  owned rights.
Reference herein to any specific commercial product, process, or service by trade name, trademark,  manufacturer,  or  otherwise  does  not  necessarily  constitute  or  imply  its endorsement,  recommendation,  or  favoring  by  the  United  States Government  or  any agency thereof.
The views and opinions of authors expressed herein do not necessarily state or reflect those of the United States Government or any agency thereof.

\end{document}